\documentclass[aps,twocolumn,pra,longbibliography,floatfix,superscriptaddress]{revtex4-1}

\usepackage{dcolumn} 
\usepackage{amsmath,amsthm,amsfonts,amssymb,bm,graphicx,color,mathpazo,times, braket}
\usepackage[colorlinks={true}, citecolor={blue}, filecolor={blue}, linkcolor={blue}, urlcolor={blue}]{hyperref}
\usepackage[caption=false]{subfig}
\usepackage{soul}
\usepackage{xcolor}

\allowdisplaybreaks

\begin{document}

\title{Classical Simulation of an All-Optical Toffoli Gate using Soliton Scattering through Asymmetric Potential Wells}

\author{Melih \"Ozkurt}
\email{mozkurt20@ku.edu.tr}
\affiliation{Department of Physics, Ko\c{c} University, 34450 Sar\i yer, Istanbul, T\"urkiye}
\author{\"Ozg\"ur E. M\"ustecapl\i o\u glu}
\email{omustecap@ku.edu.tr}
\affiliation{Department of Physics, Ko\c{c} University, 34450 Sar\i yer, Istanbul, T\"urkiye}
\affiliation{T\"UBİTAK Research Institute for Fundamental Sciences (TBAE), 41470 Gebze, T\"urkiye}

\date{\today}

\begin{abstract}
We propose and numerically simulate an all-optical Toffoli (controlled-controlled-NOT) gate based on the scattering of spatial solitons by asymmetric Pöschl-Teller potential wells. In our scheme, the logical state of the target bit is encoded in the relative spatial ordering of two distinguishable soliton components, while the control bits are represented by the presence or absence of external potential wells. We solve the nonlinear Schrödinger equations governing the soliton dynamics and systematically scan soliton amplitude and velocity, analyzing reflection and transmission coefficients to identify the operational conditions for Toffoli gate behavior. Our results demonstrate that introducing asymmetry in the potential wells significantly broadens the operational parameter window compared to symmetric configurations. We also investigate the impacts of varying degrees of asymmetry and soliton amplitude on gate performance. Furthermore, we examine the influence of weak inter-component coupling and confirm that it is not essential for gate operation. These findings generalize earlier soliton-based CNOT simulations and support the broader feasibility of classical analog modeling of multi-qubit logic gates in nonlinear optical systems.
\end{abstract}

\maketitle

\section{INTRODUCTION}
\label{sec:intro}

All-optical computing has emerged as a promising frontier in the pursuit of ultra-fast, energy-efficient information processing. Utilizing light as an information carrier offers fundamental advantages, including a wide bandwidth, ultrahigh speed, and low heat generation~\cite{tucker_role_2010}. The prospect of leveraging light to perform computation directly, not merely for transmission but also for logic and control, has stimulated active research across photonic materials, nonlinear optics, and integrated photonic circuits.

Within nonlinear optics, solitons represent a particularly attractive platform for all-optical computing. Solitons are self-reinforcing, shape-preserving wave packets resulting from a balance between nonlinear and dispersive effects~\cite{agrawal_nonlinear_2013}. Their particle-like properties make them ideal information carriers, capable of propagating over long distances without significant distortion. These properties have motivated extensive efforts to explore soliton dynamics in optical hardware, leading to the realization of various soliton-based devices~\cite{khawaja_all-optical_2016, javed_amplifying_2020, aghdami_two-dimensional_2012, Wu:07, christodoulides_blocking_2001, shaheen_adding_2020, vinayagam_stable_2018, scheuer_all-optical_2005, sabini_all-optical_1989}. These systems exploit soliton interactions either within nonlinear media or at interfaces between different nonlinear media, offering possibilities for all-optical signal processing and logical operations.

Several approaches to soliton-based logic gates have been proposed and demonstrated, employing mechanisms such as Kerr nonlinearities~\cite{scheuer_all-optical_2005, sabini_all-optical_1989}, soliton collisions~\cite{vijayajayanthi_simulation_2023, jakubowski_state_1998, maimistov_reversible_1995, anastassiou_energy-exchange_1999}, and saturable nonlinearities~\cite{peccianti_all-optical_2002, jakubowski_information_1997, sahoo_bistable_2022}. Another promising strategy is soliton scattering off external potentials~\cite{umarov_variational_2013, morales-molina_trapping_2006, al_khawaja_unidirectional_2013, al_khawaja_critical_2021, javed_unidirectional_2021, javed_unidirectional_2022, sakkaf_reflectionless_2023}, which has enabled the implementation of novel devices~\cite{krolikowski_soliton-based_1996, asad-uz-zaman_directional_2013, al_khawaja_unidirectional_2015} and can enable robust and versatile logical operations. For example, it was recently demonstrated that scattering a soliton through a carefully engineered reflectionless potential well can realize an optical analog of the controlled-NOT (CNOT) gate~\cite{javed_simulating_2022}.

In the present work, we generalize this scattering-based approach to simulate an all-optical Toffoli (controlled-controlled-NOT) gate, a three-qubit logic gate. Our scheme builds upon the classical wave analogy of quantum gates, as formulated in Ref.~\cite{javed_simulating_2022}, but aims at higher-order logic without relying on quantum entanglement or superposition. Our scheme employs two asymmetric Pöschl–Teller potential wells to encode the control qubits, where the presence or absence of a well represents a logical state. The target qubit is defined by the spatial ordering of two optical solitons belonging to different, distinguishable components. We calculate the reflection and transmission coefficients for the scattering processes to identify parameter regimes in which the soliton-potential interaction reproduces the Toffoli gate's truth table. We also investigate the effect of soliton amplitude, velocity, potential asymmetry, and weak inter-component coupling on the fidelity and robustness of gate operation.

The remainder of this paper is organized as follows: Sec.~\ref{theory_sec} introduces the theoretical model, including the nonlinear Schrödinger equations and the external potential configurations used to implement the gate dynamics. Sec.~\ref{protocol_sec} details the gate encoding scheme and the logical mapping associated with the Toffoli operation. Sec.~\ref{operation_sec} outlines the qualitative scattering behavior underlying the gate function. Sec.~\ref{num_sec} presents a detailed numerical analysis of the gate's performance, examining the effects of soliton amplitude, velocity, and potential asymmetry on the operational regime as well as cross-coupling strength. Sec.~\ref{sec:discussion} provides a discussion of the practical and foundational limitations of the scheme. We summarize and conclude in Sec.~\ref{sec:conclusion}.

\section{THEORETICAL MODEL}
\label{theory_sec}

We begin by presenting the governing equations for the dynamics of bright-bright solitons, as formulated in Refs.~\cite{javed_unidirectional_2021, javed_unidirectional_2022}. The dynamics of bright-bright solitons are governed by a system of two coupled nonlinear Schrödinger equations (CNLSEs), also known as the Manakov system~\cite{kivshar_optical_2003}:
\begin{align}
i \partial_t \psi_1 &= \left[-\frac{1}{2} \partial_x^2 - \left(g_{11} |\psi_1|^2 + g_{12} |\psi_2|^2 \right) + V(x) \right] \psi_1 \notag \\
i \partial_t \psi_2 &= \left[-\frac{1}{2} \partial_x^2 - \left(g_{12} |\psi_1|^2 + g_{22} |\psi_2|^2 \right) + V(x) \right] \psi_2
\label{cnlse}
\end{align}
where \(\psi_1(x,t)\) and \(\psi_2(x,t)\) are the complex wavefunctions of the individual soliton components, and \(V(x)\) is a common external scalar potential. The interaction type is set to attractive nonlinearity, which supports bright soliton solutions~\cite{adhikari_bright_2005}. We set the self-interaction strengths to unity, \(g_{11} = g_{22} = 1\), and assume equal cross-interactions, \(g_{12} = g_{21}\), as appropriate for the Manakov system.

In our simulations, we primarily focus on the uncoupled case \(g_{12} = g_{21} = 0\), leaving the analysis of non-zero coupling to Sec.~\ref{subsec:coupling}. The wavefunctions are initialized as exact bright soliton solutions of the homogeneous system (i.e., with \(V(x)=0\)), located far from the potential region and moving with equal velocity:
\begin{align}
\psi_1 (x,0) &= u\, \text{sech}\left[u (x - x_0)\right]\, e^{iv(x - x_0)} \notag \\
\psi_2 (x,0) &= u\, \text{sech}\left[u \left(x - (x_0 + \delta)\right)\right]\, e^{iv(x - (x_0 + \delta))}
\label{eq:initial_solitons}
\end{align}
Here, \(u\) denotes the soliton amplitude (with peak intensity proportional to \(u^2\)), and \(v\) is their shared initial velocity. The first soliton is centered at \(x_0 < 0\), and the second is offset by \(\delta\).

The external potential consists of up to two Pöschl-Teller (PT) wells~\cite{poschl_bemerkungen_1933}, modeled as:
\begin{equation}
V(x) = -\sum_{i=1}^{2} V_i\, \text{sech}^2\left(\frac{x - x_i}{w_i}\right)
\label{eq:potential}
\end{equation}
where \(V_i > 0\) is the depth, \(w_i\) is the width, and \(x_i\) is the center of the \(i\)-th well. The PT potential is known to be reflectionless in the linear regime when the condition \(V_i w_i^2 = \mathcal{N}(\mathcal{N}+1)/2\) is satisfied for integer \(\mathcal{N}\). A commonly used and practical choice is \(\mathcal{N} = 1\), which implies \(w_i = 1/\sqrt{V_i}\) and ensures reflectionless transmission for linear waves~\cite{al_khawaja_critical_2021, sakkaf_reflectionless_2023}. In the nonlinear regime, however, the potential may induce partial reflection depending on the soliton speed.

We use the split-step Fourier method (SSFM) to evolve Eqs.~\eqref{cnlse}. The simulation time is chosen to allow scattered solitons to propagate away from the potential region without re-entering due to periodic boundary conditions. We define the reflection (\(R_j\)), trapping (\(L_j\)), and transmission (\(T_j\)) coefficients for each component \(j \in \{1,2\}\) at final time \(t_f\) as:
\begin{align}
R_j &= \frac{1}{N_j(0)} \int_{-\infty}^{l_1} |\psi_j(x, t_f)|^2 \, dx \notag \\
L_j &= \frac{1}{N_j(0)} \int_{l_1}^{l_2} |\psi_j(x, t_f)|^2 \, dx \label{eq:transport} \\
T_j &= \frac{1}{N_j(0)} \int_{l_2}^{\infty} |\psi_j(x, t_f)|^2 \, dx \notag
\end{align}
The limits \(l_1\) and \(l_2\) define the region enclosing the potential(s): \(x < l_1\) captures reflected portions, \(l_1 < x < l_2\) includes trapped density, and \(x > l_2\) measures transmitted density. The normalization factor is the initial norm,
\[
N_j(0) = \int_{-\infty}^{\infty} |\psi_j(x, 0)|^2 \, dx
\]
which is conserved under Eqs.~\eqref{cnlse}.

For the uncoupled case (\(g_{12} = g_{21} = 0\)), the two components evolve independently. Since their governing equations and initial amplitudes/velocities are identical (apart from spatial offset), their scattering dynamics are equivalent. Hence, we often analyze a single generic component when characterizing the transport behavior.

\section{THE TOFFOLI GATE}
\label{protocol_sec}

The Toffoli gate, also known as the controlled-controlled-NOT (CCNOT) gate, is a three-qubit universal gate for classical reversible computation. It flips the state of a target qubit if and only if two control qubits are both in the state \(\ket{1}\). Furthermore, when combined with suitable single-qubit gates (e.g., the Hadamard gate), it forms a universal set for quantum computation. The truth table for the Toffoli gate is shown in Table~\ref{ccnot_table}.
\begin{table}[h!]
\begin{ruledtabular}
\begin{tabular}{cccc}
\multicolumn{2}{c}{\textbf{Input}} & \multicolumn{2}{c}{\textbf{Output}} \\
\cline{1-2}
\cline{3-4}
\textbf{Target} & \textbf{Control} & \textbf{Target} & \textbf{Control} \\
\hline
$\left|0\right\rangle$ & $\left|00\right\rangle$ & $\left|0\right\rangle$ & $\left|00\right\rangle$ \\
$\left|0\right\rangle$ & $\left|01\right\rangle$ & $\left|0\right\rangle$ & $\left|01\right\rangle$ \\
$\left|0\right\rangle$ & $\left|10\right\rangle$ & $\left|0\right\rangle$ & $\left|10\right\rangle$ \\
$\left|0\right\rangle$ & $\left|11\right\rangle$ & $\left|1\right\rangle$ & $\left|11\right\rangle$ \\
$\left|1\right\rangle$ & $\left|00\right\rangle$ & $\left|1\right\rangle$ & $\left|00\right\rangle$ \\
$\left|1\right\rangle$ & $\left|01\right\rangle$ & $\left|1\right\rangle$ & $\left|01\right\rangle$ \\
$\left|1\right\rangle$ & $\left|10\right\rangle$ & $\left|1\right\rangle$ & $\left|10\right\rangle$ \\
$\left|1\right\rangle$ & $\left|11\right\rangle$ & $\left|0\right\rangle$ & $\left|11\right\rangle$ \\
\end{tabular}
\end{ruledtabular}
\caption{Truth table for the Toffoli (CCNOT) gate. \(\ket{C_1 C_2}\) represents the combined state of the two control qubits, and \(\ket{T}\) is the target qubit.}
\label{ccnot_table}
\end{table}
In this work, for demonstrating the gate operation, we primarily focus on the first four input configurations where the target qubit is initially in the state \(\ket{0}\) (as listed in Table~\ref{ccnot_table}). Due to the symmetric nature of the governing equations (Eqs.~\eqref{cnlse}) for the two soliton components in the ideal case, the dynamics for the remaining four input cases (where the target qubit is initially \(\ket{1}\)) can be inferred directly. The distinction lies in the initial and final labeling of the target qubit state, which is defined by the relative spatial ordering of the two solitons. Therefore, analyzing the first four configurations is sufficient to characterize the fundamental soliton dynamics that implement the Toffoli gate's logic.

\section{OPERATIONAL PRINCIPLE OF THE SOLITON-BASED TOFFOLI GATE}
\label{operation_sec}

In our proposed scheme, the target qubit is encoded in the spatial ordering of two distinguishable solitons with equal amplitudes but belonging to different components. Specifically, the target qubit (\(\left|T\right\rangle\))  is defined as being in the \(\left|0\right\rangle\) state when the soliton from the first component is to the right of the soliton from the second component, and in the \(\left|1\right\rangle\) state otherwise. This encoding uses the relative positions of the solitons rather than their individual properties.

For the control qubits, we adopt a scheme inspired by Ref.~\cite{javed_simulating_2022}, in which the presence or absence of localized Pöschl–Teller potential wells encodes logical states. The first control qubit is set to \(\left|1\right\rangle\) if a potential well is placed at position \(x_1\) and to \(\left|0\right\rangle\) otherwise. Similarly, the second control qubit is set to \(\left|1\right\rangle\)  if a well is placed at \(x_2\) and to \(\left|0\right\rangle\) otherwise. This yields the four standard control states: \(\left|00\right\rangle\), \(\left|01\right\rangle\), \(\left|10\right\rangle\), and \(\left|11\right\rangle\), which gives four different scenarios for our analysis. 

\begin{figure}[tb]
\centering
\includegraphics[width=0.9\linewidth]{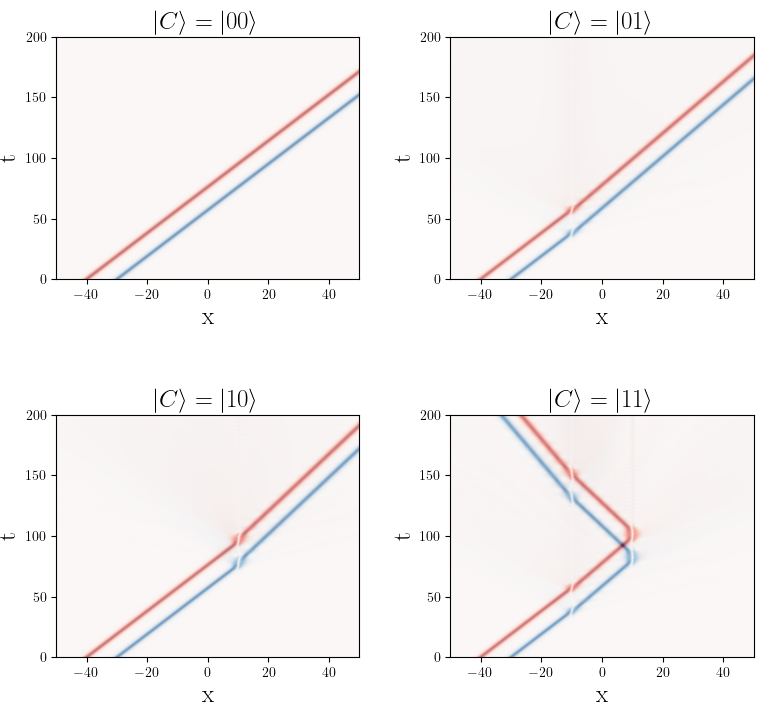}
\caption{Space-time diagrams illustrating the Toffoli gate operation for four control configurations \(\ket{C_1 C_2}\), when the target qubit is set to \(\ket{0}\).  The blue soliton represents component ($\psi_1$), and the red soliton represents component ($\psi_2$). The target state flips to \(\ket{1}\)) (i.e., solitons exchange positions) only in the \(\ket{11}\) control case, confirming the conditional behavior of the CCNOT gate. Parameters: soliton amplitude \(u=1.4\), velocity $v = 0.525$, initial center of \(\psi_1\) at \(x_0=-30\), separation \(\delta=-10\) (so \(\psi_2\) is at \(x=-40\)). For the \(\ket{11}\) case, potential well 1 has depth \(V_1=4.32\) centered at \(x_1=-10\), and potential well 2 has depth \(V_2=4\) centered at \(x_2=10\). Both wells have width \(w_1=w_2=0.5\). For \(\ket{10}\), only well $1$ is present; for \(\ket{01}\), only well $2$ is present.}
\label{fig:spacetime}
\end{figure}

Simulation results demonstrate the core mechanism of the gate. When only one potential well is present, (control states 
\(\left|C\right\rangle = \left|01\right\rangle\) or \(\left|10\right\rangle\)) solitons pass through with negligible disturbance, and the original ordering, and thus the target qubit state, is preserved. However, in the (\(\left|C\right\rangle = \left|11\right\rangle\)) case, where both potential wells are present, the soliton closer to the interaction region is reflected, causing the solitons to switch order and the target state is flipped. These scenarios are illustrated in the space-time diagrams of Fig.~\ref{fig:spacetime}, confirming the Toffoli gate functionality.

For control states $\ket{00}$, $\ket{01}$, and $\ket{10}$, the solitons propagate without altering their initial relative ordering, indicating no change in the target qubit state. Specifically, when only one potential well is present ($\ket{C} = \ket{01}$ or $\ket{10}$), solitons undergo complete transmission, thereby preserving their initial ordering. In contrast, for the $\ket{11}$ control state, where both potential wells are present, total reflection occurs. Initially, $\psi_1$ (blue) is placed right of $\psi_2$ (red); however, interaction with the wells reverses their order, flipping the target qubit state as required by the CCNOT gate operation.

\begin{figure}[thb]
\includegraphics[width=0.9\linewidth]{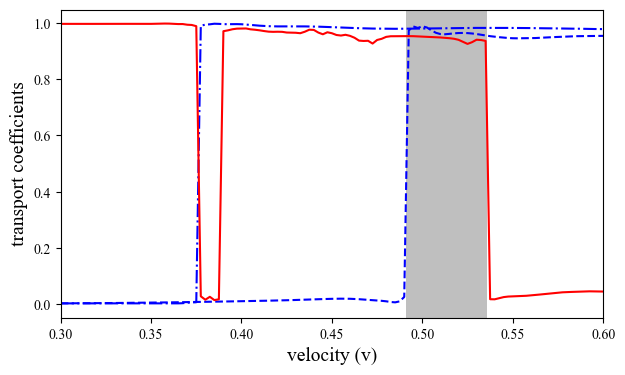}
\caption{Transport coefficients for a single soliton component as a function of its initial velocity $v$. The red solid line shows the reflectance $R_{11}$ for the double-well configuration (control state \(\ket{11}\)). The dashed and dot-dashed blue lines show the transmittances $T_{01}$ and $T_{10}$ for configurations with only the second (at $x_2=10$, control state \(\ket{01}\)) or first (at $x_1=-10$, control state \(\ket{10}\)) well present, respectively. The gray shaded region marks the operational velocity window for the Toffoli gate, where high reflection occurs only in the
\(\ket{11}\)) case. Parameters: amplitude $u = 1.4$. For the \(\ket{11}\) case, $V_1=4.32$ (depth of well at $x_1=-10$) and $V_2=4$ (depth of well at $x_2=10$). For \(\ket{10}\), only the well at $x_1=-10$ with depth $V_1=4.32$ is present. For \(\ket{01}\), only the well at $x_2=10$ with depth $V_2=4$ is present.}
\label{fig:transport_coeff1D}
\end{figure}

As discussed in Sec.~\ref{theory_sec}, due to the uncoupled nature of Eqs.~\eqref{cnlse} in the ideal case without cross coupling coefficients, we analyze the transport coefficients for a single soliton component. The operational window is defined by the range of velocities $v$ where solitons undergo strong transmission in single-well configurations ($T_{01} \approx 1$, $T_{10} \approx 1$) and strong reflection in the double-well configuration ($R_{11} \approx 1$). Figure~\ref{fig:transport_coeff1D} presents the transport coefficients as a function of soliton velocity. This velocity window, shaded in gray, lies approximately within \(0.4925 \le v \le 0.5350\) for the parameters used. Within this range, we find:

\begin{itemize}
    \item For the double-well case (\(\ket{11}\)), the reflectance is high, with \(R_{11} \ge 0.94\).
    \item For the single-well cases:
    \begin{itemize}
        \item When only the well at \(x_1\) is present (\(\ket{10}\)), transmission is high with \(T_{10} \ge 0.97\).
        \item When only the well at \(x_2\) is present (\(\ket{01}\)), transmission is high with \(T_{01} \ge 0.98\).
    \end{itemize}
\end{itemize}

The gate mechanism harnesses the velocity dependence of soliton interactions with potential wells. The critical velocity ($v_c$) is the minimum incident velocity required for a soliton to predominantly transmit through a given potential structure. The underlying principle of the CCNOT gate operation involves tailoring the critical velocities for the two potential wells and exploiting the velocity reduction a soliton experiences upon interacting with a potential well~\cite{asad-uz-zaman_directional_2013}. 

Let's focus on tailoring this velocity selective reflection/transmission behavior for the \(\ket{11}\) case, which is key to the gate's operation. In the \(\ket{11}\) configuration (wells at \(x_1\) and \(x_2\), with \(x_1 < x_2\)), an incident soliton first encounters the well at \(x_1\). This interaction reduces the soliton's velocity. If the soliton (with its reduced velocity) then encounters the second well at \(x_2\) and its velocity is below \(v_{c2}\) (the critical velocity for transmission through the second well), it will be reflected by the well at \(x_2\). For this reflected soliton to exit the two-well system in the backward direction (i.e., to achieve total reflection from the combined \(V(x_1) + V(x_2)\) structure), it must then propagate back toward the first well at \(x_1\). If its velocity upon reaching \(x_1\) (from the right) is sufficient to transmit through \(x_1\) in the reverse direction, then overall reflection from the double-well system is achieved.

The asymmetry in the potential wells (i.e., \(V_1 \ne V_2\)) is crucial for creating a velocity window where this sequence occurs effectively. The precise conditions for this behavior can be tuned by adjusting the depths of the potential wells, as the depth significantly influences the critical velocity (deeper wells typically result in lower critical velocities). The following section numerically investigates how the relative depths of the potential wells (i.e., their asymmetry) and the amplitude of the incident soliton influence these operational characteristics.

\section{NUMERICAL RESULTS}
\label{num_sec}

\subsection{Effect of Potential Asymmetry on Gate Operation}
\label{subsec:asymmetry}

\begin{figure}[b]
\centering
\begin{minipage}{0.48\linewidth}
  \includegraphics[width=\linewidth]{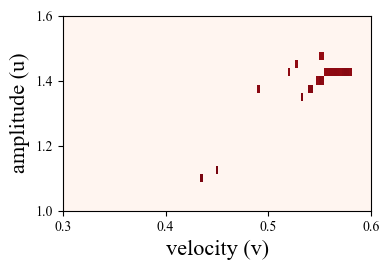}
  \centerline{\textbf{(a) $\alpha = 1.00$}}
\end{minipage}\hfill
\begin{minipage}{0.48\linewidth}
  \includegraphics[width=\linewidth]{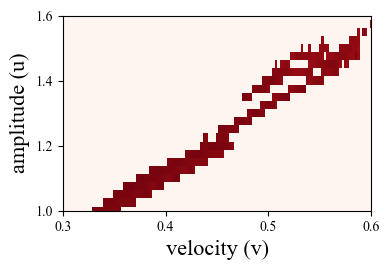}
  \centerline{\textbf{(b) $\alpha = 1.04$}}
\end{minipage}
\vspace{0.5em}
\begin{minipage}{0.48\linewidth}
  \includegraphics[width=\linewidth]{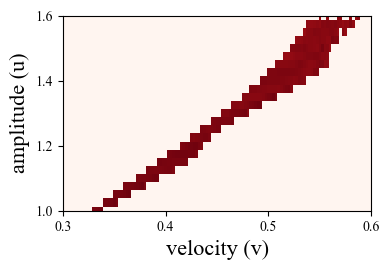}
  \centerline{\textbf{(c) $\alpha = 1.08$}}
\end{minipage}\hfill
\begin{minipage}{0.48\linewidth}
  \includegraphics[width=\linewidth]{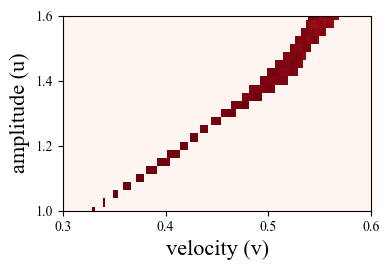}
  \centerline{\textbf{(d) $\alpha = 1.12$}}
\end{minipage}
\caption{Operational regions for the Toffoli gate in the $(v,u)$ plane, for different asymmetry parameters $\alpha$. The shaded areas represent parameter combinations satisfying $R_{11} > 0.9$ (for the \(\ket{11}\) state), $T_{10} > 0.9$ (for the \(\ket{10}\) state, with well $V_1=\alpha V_0$), and $T_{01} > 0.9$ (for the \(\ket{01}\) state, with well $V_2=V_0=4$).}
\label{fig:op_regions_alpha}
\end{figure}

We now investigate the effect of potential asymmetry on the Toffoli gate's operational parameters, specifically the incident soliton velocity \(v\) and amplitude \(u\). Asymmetry is introduced by varying the depth of the first potential well \(V_1 = \alpha V_0\), while keeping the second well depth fixed at \(V_2 = V_0\) (with \(V_0 = 4\)). Other parameters are held constant: well centers at \(x_1 = -10\) and \(x_2 = 10\), and widths \(w_1 = w_2 = 0.5\). For these parameters, the product \(V_0 w_0^2 = 1\), corresponding to a first-order reflectionless PT potential in the linear regime. 

The operational region for the Toffoli gate is defined by the set of parameter pairs \((v, u)\) that simultaneously satisfy three conditions: high reflectance \((R_{11} > 0.9)\) for the double-well configuration (control state \(\ket{11}\)), and high transmittance \((T_{10} > 0.9\) and \(T_{01} > 0.9)\) through the individual wells \(V_1 = \alpha V_0\) and \(V_2 = V_0\), respectively (control states \(\ket{10}\) and \(\ket{01}\)). Figure~\ref{fig:op_regions_alpha} maps these operational regions for various values of \(\alpha\).

The results in Fig.~\ref{fig:op_regions_alpha} reveal that introducing asymmetry \((\alpha \neq 1.0)\) significantly impacts the operational window. Increasing the asymmetry parameter \(\alpha\) initially broadens this region by shifting the critical velocities in a favorable way. However, if \(\alpha\) becomes too large, the soliton experiences less velocity reduction after the first well, resulting in a velocity that may exceed the critical speed of the second well, thereby reducing the usable parameter space. Therefore, there exists an optimal balance in choosing \(\alpha\). For instance, for soliton amplitudes \(1.0 < u < 1.2\), \(\alpha \approx 1.04\) yields the widest operational window. For higher amplitudes, such as \(1.2 < u < 1.6\), \(\alpha \approx 1.08\) is more effective. It is noteworthy that symmetric potentials \((\alpha = 1.0)\) also allow for Toffoli gate operation, though typically with narrower and less robust operational windows compared to appropriately chosen asymmetric configurations.

\subsection{Effect of Cross-Component Coupling on Gate Operation}
\label{subsec:coupling}

\begin{figure}[tb]
\centering
\begin{minipage}{0.48\linewidth}
\includegraphics[width=\linewidth]{Figures/analysis/analysis1.08.png} 
\centerline{\textbf{(a) $g_{12} = 0$}}
\end{minipage}\hfill
\begin{minipage}{0.48\linewidth}
\includegraphics[width=\linewidth]{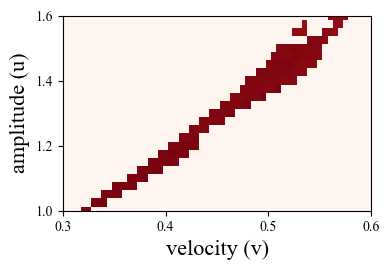}
\centerline{\textbf{(b) $g_{12} = 0.1$}}
\end{minipage}
\vspace{0.5em}
\begin{minipage}{0.48\linewidth}
\includegraphics[width=\linewidth]{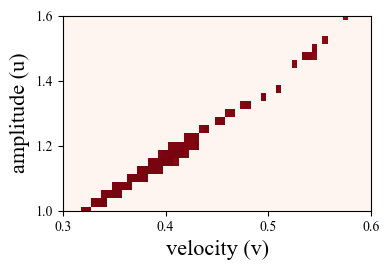}
\centerline{\textbf{(c) $g_{12} = 0.2$}}
\end{minipage}\hfill
\begin{minipage}{0.48\linewidth}
\includegraphics[width=\linewidth]{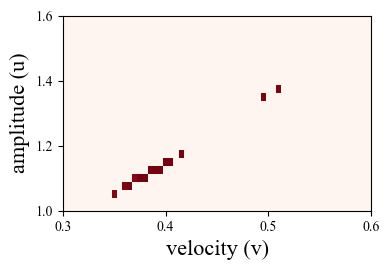}
\centerline{\textbf{(d) $g_{12} = 0.3$}}
\end{minipage}
\caption{Operational regions for the Toffoli gate with asymmetry $\alpha=1.08$, for varying inter-component coupling strengths $g_{12} (=g_{21})$. Panel (a) ($g_{12}=0$) is reproduced from Fig.~\ref{fig:op_regions_alpha}(c) for comparison.}
\label{fig:op_regions_coupling}
\end{figure}

We now examine a more realistic setting by incorporating a nonzero cross-component coupling term ($g_{12} = g_{21} \neq 0$) in the coupled nonlinear Schrödinger equations (CNLSEs, Eqs.~\eqref{cnlse}), with intra-component interactions fixed at $g_{11} = g_{22} = 1$. Motivated by the favorable performance observed in the uncoupled case, we fix the asymmetry parameter at $\alpha = 1.08$ (i.e., $V_1 = 1.08 V_0$, $V_2 = V_0$, with $V_0 = 4$) and study the effect of varying the cross-coupling strength $g_{12}$ over the range $0 \leq g_{12} \leq 0.3$.

Figure~\ref{fig:op_regions_coupling} shows the resulting operational regions for $g_{12} = 0$, $0.1$, $0.2$, and $0.3$. As seen by comparing panels (a)–(d), increasing $g_{12}$ progressively reduces and distorts the operational window, especially for solitons with higher amplitudes $u$. This effect is attributable to increased nonlinear interactions between the components, which perturb the scattering dynamics and compromise the conditional transmission/reflection behavior required for gate operation. For $g_{12} \gtrsim 0.3$, no reliable operational region is observed under the current parameter settings, indicating that strong cross-component coupling undermines the gate’s selectivity. These findings suggest that weak inter-component coupling is preferable for stable and high-fidelity Toffoli gate implementation in this platform.

\section{DISCUSSION}
\label{sec:discussion}

Despite these promising numerical results, experimental implementation faces several practical and fundamental challenges. Chief among them is the difficulty of dynamically implementing and modulating potential wells ($V_1$, $V_2$) in a real optical medium. One potential solution is to design a set of photonic materials or metastructures with pre-engineered potentials and then route solitons through them based on the logical state of control bits using fast optical switches or shunt networks~\cite{Wu:07}. Another possibility is the use of externally reconfigurable optical potentials, such as active reconfigurable optical metasurfaces~\cite{Neshev2018}, for example, using thermo-optic, electro-optic, or all-optical modulation techniques, which could allow real-time control over gate behavior.

Another limitation is the non-reversibility of this gate implementation, which contrasts with the theoretically reversible nature of classical Toffoli logic. However, it remains an open question whether practical computation with soliton-based gates requires strict reversibility. In conventional quantum computation, reversibility is a formal requirement due to unitarity, but real devices are typically open systems subject to decoherence and loss, thereby breaking strict reversibility. In optical logic and photonic computing, many practical schemes—especially those involving dissipative or nonlinear processes—rely on one-way transformations, which may still be sufficient for efficient classical or hybrid information processing.

There are two technical points that can influence the preparation of the optical solitons in our scheme. First, we assume equal amplitude but distinguishable solitons, and second, precise preparation of their initial speeds. The distinguishability of solitons can be physically realized using orthogonal polarization states or distinct carrier frequencies. In typical nonlinear optical media, the Pöschl–Teller-type potential can be engineered to affect both components similarly, provided the medium supports two weakly interacting modes and is not strongly birefringent. This ensures the physical implementability of the control and target bit encoding scheme used in this work.
The operational velocity windows for the three-qubit Toffoli gate tend to be somewhat narrower than those reported for the two-qubit CNOT gate in Ref.~\cite{javed_simulating_2022}. This is likely attributable to the increased complexity and stricter conditions required for the three-input, three-output logic of the Toffoli gate, involving interactions with up to two potential wells. Regarding the effects of cross-component coupling ($g_{12}$), while the gate can tolerate small cross-coupling strengths, increasing $g_{12}$ significantly degrades performance by shrinking and distorting the operational region, particularly for higher soliton amplitudes. 

There are also fundamental shortcomings that need to be discussed. Using classical systems, such as soliton dynamics, for simulating the logical behavior (truth table) of a quantum gate, does not encompass the full scope of the quantum gate functionality, due to the absence of genuine quantum features, like handling quantum superpositions and entanglement. The value of classical simulation of quantum gates stems from the ability to visualize and validate the logic structure of quantum circuits without 
requiring a full quantum setup. Classical simulators can serve for benchmarking and early-prototyping as well.
In addition, even without genuine quantum effects, such simulations might 
inspire high-speed and low-power architectures for nonlinear optical computing, future integrated 
photonic circuits, and possibly including hybrid quantum-classical systems.
 
\section{CONCLUSION}
\label{sec:conclusion}

In summary, we have numerically demonstrated a scheme for realizing an all-optical Toffoli gate based on the scattering of solitons by asymmetric Pöschl–Teller potential wells. The target bit is encoded in the relative spatial ordering of two solitons belonging to distinct, distinguishable components, while the two control bits are encoded by the presence or absence of these potential wells. A three-qubit Toffoli gate, or a CCNOT gate, is a natural generalization of existing classically simulated CNOT gates. While CNOT gates are sufficient for universality when paired with single-qubit gates, Toffoli gates are natural generalizations that help simulate more complex circuits or error correction routines. 

We systematically investigated the influence of potential well asymmetry (parameterized by $\alpha$) and incident soliton parameters (amplitude $u$ and velocity $v$) on the gate's operational conditions. This involved identifying parameter regimes that ensure high reflection for the \(\ket{11}\) control state (both wells present) and high transmission for the \(\ket{01}\) and \(\ket{10}\) control states (single well present). A key objective of this numerical analysis was to identify parameter configurations that offer robust gate operation against potential fluctuations in soliton amplitude and velocity. Our findings indicate that introducing an appropriate degree of asymmetry in the potential wells ($\alpha \neq 1$) can significantly broaden the operational velocity window compared to the symmetric case ($\alpha=1$). The operational window's position and width in the $(v,u)$ plane are sensitive to the soliton amplitude $u$. Adjusting the input soliton amplitude allows the operational velocity range to be shifted, a tunability also noted for CNOT gates in Ref.~\cite{javed_simulating_2022}. This tunability is a potentially valuable feature for practical implementations. However, this tuning does not necessarily preserve the width of the operational window; certain amplitude ranges yield broader and more robust operational characteristics than others. Overall, as shown in Fig.~\ref{fig:op_regions_alpha}, by carefully selecting the asymmetry and soliton amplitude, the proposed Toffoli gate can achieve reliable performance across a notable range of soliton velocities.

Our results contribute to the development of all-optical logic gates, which are essential for ultra-fast and low-power optical computing systems. Simulating quantum logic operations using classical systems can be valuable for visualizing and testing concepts and gate behaviors in a more accessible setting or can aid the development of hybrid classical-quantum computing architectures. 

\newpage 
\bibliography{references}

\begin{thebibliography}{33}%
\makeatletter
\providecommand \@ifxundefined [1]{%
 \@ifx{#1\undefined}
}%
\providecommand \@ifnum [1]{%
 \ifnum #1\expandafter \@firstoftwo
 \else \expandafter \@secondoftwo
 \fi
}%
\providecommand \@ifx [1]{%
 \ifx #1\expandafter \@firstoftwo
 \else \expandafter \@secondoftwo
 \fi
}%
\providecommand \natexlab [1]{#1}%
\providecommand \enquote  [1]{``#1''}%
\providecommand \bibnamefont  [1]{#1}%
\providecommand \bibfnamefont [1]{#1}%
\providecommand \citenamefont [1]{#1}%
\providecommand \href@noop [0]{\@secondoftwo}%
\providecommand \href [0]{\begingroup \@sanitize@url \@href}%
\providecommand \@href[1]{\@@startlink{#1}\@@href}%
\providecommand \@@href[1]{\endgroup#1\@@endlink}%
\providecommand \@sanitize@url [0]{\catcode `\\12\catcode `\$12\catcode `\&12\catcode `\#12\catcode `\^12\catcode `\_12\catcode `\%12\relax}%
\providecommand \@@startlink[1]{}%
\providecommand \@@endlink[0]{}%
\providecommand \url  [0]{\begingroup\@sanitize@url \@url }%
\providecommand \@url [1]{\endgroup\@href {#1}{\urlprefix }}%
\providecommand \urlprefix  [0]{URL }%
\providecommand \Eprint [0]{\href }%
\providecommand \doibase [0]{http://dx.doi.org/}%
\providecommand \selectlanguage [0]{\@gobble}%
\providecommand \bibinfo  [0]{\@secondoftwo}%
\providecommand \bibfield  [0]{\@secondoftwo}%
\providecommand \translation [1]{[#1]}%
\providecommand \BibitemOpen [0]{}%
\providecommand \bibitemStop [0]{}%
\providecommand \bibitemNoStop [0]{.\EOS\space}%
\providecommand \EOS [0]{\spacefactor3000\relax}%
\providecommand \BibitemShut  [1]{\csname bibitem#1\endcsname}%
\let\auto@bib@innerbib\@empty
\bibitem [{\citenamefont {Tucker}(2010)}]{tucker_role_2010}%
  \BibitemOpen
  \bibfield  {author} {\bibinfo {author} {\bibfnamefont {Rodney~S.}\ \bibnamefont {Tucker}},\ }\bibfield  {title} {\enquote {\bibinfo {title} {The role of optics in computing},}\ }\href {\doibase 10.1038/nphoton.2010.162} {\bibfield  {journal} {\bibinfo  {journal} {Nature Photonics}\ }\textbf {\bibinfo {volume} {4}},\ \bibinfo {pages} {405--405} (\bibinfo {year} {2010})}\BibitemShut {NoStop}%
\bibitem [{\citenamefont {Agrawal}(2013)}]{agrawal_nonlinear_2013}%
  \BibitemOpen
  \bibfield  {author} {\bibinfo {author} {\bibfnamefont {G.~P.}\ \bibnamefont {Agrawal}},\ }\href@noop {} {\emph {\bibinfo {title} {Nonlinear {Fiber} {Optics}}}},\ Optics and {Photonics}\ (\bibinfo  {publisher} {Elsevier Science},\ \bibinfo {year} {2013})\BibitemShut {NoStop}%
\bibitem [{\citenamefont {Khawaja}\ \emph {et~al.}(2016)\citenamefont {Khawaja}, \citenamefont {Al-Marzoug},\ and\ \citenamefont {Bahlouli}}]{khawaja_all-optical_2016}%
  \BibitemOpen
  \bibfield  {author} {\bibinfo {author} {\bibfnamefont {U.~Al}\ \bibnamefont {Khawaja}}, \bibinfo {author} {\bibfnamefont {S.~M.}\ \bibnamefont {Al-Marzoug}}, \ and\ \bibinfo {author} {\bibfnamefont {H.}~\bibnamefont {Bahlouli}},\ }\bibfield  {title} {\enquote {\bibinfo {title} {All-optical switches, unidirectional flow, and logic gates with discrete solitons in waveguide arrays},}\ }\href {\doibase 10.1364/OE.24.011062} {\bibfield  {journal} {\bibinfo  {journal} {Optics Express}\ }\textbf {\bibinfo {volume} {24}},\ \bibinfo {pages} {11062--11074} (\bibinfo {year} {2016})}\BibitemShut {NoStop}%
\bibitem [{\citenamefont {Javed}\ \emph {et~al.}(2020)\citenamefont {Javed}, \citenamefont {Shaheen},\ and\ \citenamefont {Al~Khawaja}}]{javed_amplifying_2020}%
  \BibitemOpen
  \bibfield  {author} {\bibinfo {author} {\bibfnamefont {Amaria}\ \bibnamefont {Javed}}, \bibinfo {author} {\bibfnamefont {Alaa}\ \bibnamefont {Shaheen}}, \ and\ \bibinfo {author} {\bibfnamefont {U.}~\bibnamefont {Al~Khawaja}},\ }\bibfield  {title} {\enquote {\bibinfo {title} {Amplifying optical signals with discrete solitons in waveguide arrays},}\ }\href {\doibase 10.1016/j.physleta.2020.126654} {\bibfield  {journal} {\bibinfo  {journal} {Physics Letters A}\ }\textbf {\bibinfo {volume} {384}},\ \bibinfo {pages} {126654} (\bibinfo {year} {2020})}\BibitemShut {NoStop}%
\bibitem [{\citenamefont {Aghdami}\ \emph {et~al.}(2012)\citenamefont {Aghdami}, \citenamefont {Golshani},\ and\ \citenamefont {Kheradmand}}]{aghdami_two-dimensional_2012}%
  \BibitemOpen
  \bibfield  {author} {\bibinfo {author} {\bibfnamefont {K.~M.}\ \bibnamefont {Aghdami}}, \bibinfo {author} {\bibfnamefont {M.}~\bibnamefont {Golshani}}, \ and\ \bibinfo {author} {\bibfnamefont {R.}~\bibnamefont {Kheradmand}},\ }\bibfield  {title} {\enquote {\bibinfo {title} {Two-{Dimensional} {Discrete} {Cavity} {Solitons}: {Switching} and {All}-{Optical} {Gates}},}\ }\href {\doibase 10.1109/JPHOT.2012.2205234} {\bibfield  {journal} {\bibinfo  {journal} {IEEE Photonics Journal}\ }\textbf {\bibinfo {volume} {4}},\ \bibinfo {pages} {1147--1154} (\bibinfo {year} {2012})}\BibitemShut {NoStop}%
\bibitem [{\citenamefont {Wu}\ \emph {et~al.}(2007)\citenamefont {Wu}, \citenamefont {Huang}, \citenamefont {Chen},\ and\ \citenamefont {Tasy}}]{Wu:07}%
  \BibitemOpen
  \bibfield  {author} {\bibinfo {author} {\bibfnamefont {Yaw-Dong}\ \bibnamefont {Wu}}, \bibinfo {author} {\bibfnamefont {Min-Lun}\ \bibnamefont {Huang}}, \bibinfo {author} {\bibfnamefont {Mao-Hsiung}\ \bibnamefont {Chen}}, \ and\ \bibinfo {author} {\bibfnamefont {Rong-Zhan}\ \bibnamefont {Tasy}},\ }\bibfield  {title} {\enquote {\bibinfo {title} {All-optical switch based on the local nonlinear mach-zehnder interferometer},}\ }\href {\doibase 10.1364/OE.15.009883} {\bibfield  {journal} {\bibinfo  {journal} {Opt. Express}\ }\textbf {\bibinfo {volume} {15}},\ \bibinfo {pages} {9883--9892} (\bibinfo {year} {2007})}\BibitemShut {NoStop}%
\bibitem [{\citenamefont {Christodoulides}\ and\ \citenamefont {Eugenieva}(2001)}]{christodoulides_blocking_2001}%
  \BibitemOpen
  \bibfield  {author} {\bibinfo {author} {\bibfnamefont {D.N.}\ \bibnamefont {Christodoulides}}\ and\ \bibinfo {author} {\bibfnamefont {E.D.}\ \bibnamefont {Eugenieva}},\ }\bibfield  {title} {\enquote {\bibinfo {title} {Blocking and routing discrete solitons in two-dimensional networks of nonlinear waveguide arrays},}\ }\href {\doibase 10.1103/PhysRevLett.87.233901} {\bibfield  {journal} {\bibinfo  {journal} {Physical Review Letters}\ }\textbf {\bibinfo {volume} {87}},\ \bibinfo {pages} {233901} (\bibinfo {year} {2001})}\BibitemShut {NoStop}%
\bibitem [{\citenamefont {Shaheen}\ \emph {et~al.}(2020)\citenamefont {Shaheen}, \citenamefont {Javed},\ and\ \citenamefont {Khawaja}}]{shaheen_adding_2020}%
  \BibitemOpen
  \bibfield  {author} {\bibinfo {author} {\bibfnamefont {Alaa}\ \bibnamefont {Shaheen}}, \bibinfo {author} {\bibfnamefont {Amaria}\ \bibnamefont {Javed}}, \ and\ \bibinfo {author} {\bibfnamefont {U~Al}\ \bibnamefont {Khawaja}},\ }\bibfield  {title} {\enquote {\bibinfo {title} {Adding binary numbers with discrete solitons in waveguide arrays},}\ }\href {\doibase 10.1088/1402-4896/aba2b2} {\bibfield  {journal} {\bibinfo  {journal} {Physica Scripta}\ }\textbf {\bibinfo {volume} {95}},\ \bibinfo {pages} {085107} (\bibinfo {year} {2020})}\BibitemShut {NoStop}%
\bibitem [{\citenamefont {Vinayagam}\ \emph {et~al.}(2018)\citenamefont {Vinayagam}, \citenamefont {Javed},\ and\ \citenamefont {Al~Khawaja}}]{vinayagam_stable_2018}%
  \BibitemOpen
  \bibfield  {author} {\bibinfo {author} {\bibfnamefont {P.S.}\ \bibnamefont {Vinayagam}}, \bibinfo {author} {\bibfnamefont {A.}~\bibnamefont {Javed}}, \ and\ \bibinfo {author} {\bibfnamefont {U.}~\bibnamefont {Al~Khawaja}},\ }\bibfield  {title} {\enquote {\bibinfo {title} {Stable discrete soliton molecules in two-dimensional waveguide arrays},}\ }\href {\doibase 10.1103/PhysRevA.98.063839} {\bibfield  {journal} {\bibinfo  {journal} {Physical Review A}\ }\textbf {\bibinfo {volume} {98}},\ \bibinfo {pages} {063839} (\bibinfo {year} {2018})}\BibitemShut {NoStop}%
\bibitem [{\citenamefont {Scheuer}\ and\ \citenamefont {Orenstein}(2005)}]{scheuer_all-optical_2005}%
  \BibitemOpen
  \bibfield  {author} {\bibinfo {author} {\bibfnamefont {Jacob}\ \bibnamefont {Scheuer}}\ and\ \bibinfo {author} {\bibfnamefont {Meir}\ \bibnamefont {Orenstein}},\ }\bibfield  {title} {\enquote {\bibinfo {title} {All-optical sequential logic using coupled-resonators-based optical switches},}\ }\href@noop {} {\bibfield  {journal} {\bibinfo  {journal} {Journal of the Optical Society of America B}\ }\textbf {\bibinfo {volume} {22}},\ \bibinfo {pages} {1053} (\bibinfo {year} {2005})}\BibitemShut {NoStop}%
\bibitem [{\citenamefont {Sabini}\ \emph {et~al.}(1989)\citenamefont {Sabini}, \citenamefont {Finlayson},\ and\ \citenamefont {Stegeman}}]{sabini_all-optical_1989}%
  \BibitemOpen
  \bibfield  {author} {\bibinfo {author} {\bibfnamefont {J.P.}\ \bibnamefont {Sabini}}, \bibinfo {author} {\bibfnamefont {N.}~\bibnamefont {Finlayson}}, \ and\ \bibinfo {author} {\bibfnamefont {G.I.}\ \bibnamefont {Stegeman}},\ }\bibfield  {title} {\enquote {\bibinfo {title} {All-optical switching in nonlinear {X} junctions},}\ }\href {\doibase 10.1063/1.101689} {\bibfield  {journal} {\bibinfo  {journal} {Applied Physics Letters}\ }\textbf {\bibinfo {volume} {55}},\ \bibinfo {pages} {1176--1178} (\bibinfo {year} {1989})}\BibitemShut {NoStop}%
\bibitem [{\citenamefont {Vijayajayanthi}\ \emph {et~al.}(2023)\citenamefont {Vijayajayanthi}, \citenamefont {Kanna},\ and\ \citenamefont {Lakshmanan}}]{vijayajayanthi_simulation_2023}%
  \BibitemOpen
  \bibfield  {author} {\bibinfo {author} {\bibfnamefont {M.}~\bibnamefont {Vijayajayanthi}}, \bibinfo {author} {\bibfnamefont {T.}~\bibnamefont {Kanna}}, \ and\ \bibinfo {author} {\bibfnamefont {M.}~\bibnamefont {Lakshmanan}},\ }\bibfield  {title} {\enquote {\bibinfo {title} {Simulation of universal optical logic gates under energy sharing collisions of {Manakov} solitons and fulfillment of practical optical logic criteria},}\ }\href {\doibase 10.1103/PhysRevE.108.054213} {\bibfield  {journal} {\bibinfo  {journal} {Physical Review E}\ }\textbf {\bibinfo {volume} {108}},\ \bibinfo {pages} {054213} (\bibinfo {year} {2023})}\BibitemShut {NoStop}%
\bibitem [{\citenamefont {Jakubowski}\ \emph {et~al.}(1998)\citenamefont {Jakubowski}, \citenamefont {Steiglitz},\ and\ \citenamefont {Squier}}]{jakubowski_state_1998}%
  \BibitemOpen
  \bibfield  {author} {\bibinfo {author} {\bibfnamefont {Mariusz~H.}\ \bibnamefont {Jakubowski}}, \bibinfo {author} {\bibfnamefont {Ken}\ \bibnamefont {Steiglitz}}, \ and\ \bibinfo {author} {\bibfnamefont {Richard}\ \bibnamefont {Squier}},\ }\bibfield  {title} {\enquote {\bibinfo {title} {State transformations of colliding optical solitons and possible application to computation in bulk media},}\ }\href {\doibase 10.1103/PhysRevE.58.6752} {\bibfield  {journal} {\bibinfo  {journal} {Physical Review E}\ }\textbf {\bibinfo {volume} {58}},\ \bibinfo {pages} {6752--6758} (\bibinfo {year} {1998})}\BibitemShut {NoStop}%
\bibitem [{\citenamefont {Maimistov}(1995)}]{maimistov_reversible_1995}%
  \BibitemOpen
  \bibfield  {author} {\bibinfo {author} {\bibfnamefont {Andrei~I.}\ \bibnamefont {Maimistov}},\ }\bibfield  {title} {\enquote {\bibinfo {title} {Reversible logic elements as a new field of application of optical solitons},}\ }\href {\doibase 10.1070/QE1995v025n10ABEH000520} {\bibfield  {journal} {\bibinfo  {journal} {Quantum Electronics}\ }\textbf {\bibinfo {volume} {25}},\ \bibinfo {pages} {1009} (\bibinfo {year} {1995})}\BibitemShut {NoStop}%
\bibitem [{\citenamefont {Anastassiou}\ \emph {et~al.}(1999)\citenamefont {Anastassiou}, \citenamefont {Segev}, \citenamefont {Steiglitz}, \citenamefont {Giordmaine}, \citenamefont {Mitchell}, \citenamefont {Shih}, \citenamefont {Lan},\ and\ \citenamefont {Martin}}]{anastassiou_energy-exchange_1999}%
  \BibitemOpen
  \bibfield  {author} {\bibinfo {author} {\bibfnamefont {Charalambos}\ \bibnamefont {Anastassiou}}, \bibinfo {author} {\bibfnamefont {Mordechai}\ \bibnamefont {Segev}}, \bibinfo {author} {\bibfnamefont {Kenneth}\ \bibnamefont {Steiglitz}}, \bibinfo {author} {\bibfnamefont {JA}~\bibnamefont {Giordmaine}}, \bibinfo {author} {\bibfnamefont {Matthew}\ \bibnamefont {Mitchell}}, \bibinfo {author} {\bibfnamefont {Ming-Feng}\ \bibnamefont {Shih}}, \bibinfo {author} {\bibfnamefont {Song}\ \bibnamefont {Lan}}, \ and\ \bibinfo {author} {\bibfnamefont {John}\ \bibnamefont {Martin}},\ }\bibfield  {title} {\enquote {\bibinfo {title} {Energy-{Exchange} {Interactions} between {Colliding} {Vector} {Solitons}},}\ }\href {\doibase 10.1103/PhysRevLett.83.2332} {\bibfield  {journal} {\bibinfo  {journal} {Physical Review Letters}\ }\textbf {\bibinfo {volume} {83}},\ \bibinfo {pages} {2332--2335} (\bibinfo {year} {1999})}\BibitemShut {NoStop}%
\bibitem [{\citenamefont {Peccianti}\ \emph {et~al.}(2002)\citenamefont {Peccianti}, \citenamefont {Conti}, \citenamefont {Assanto}, \citenamefont {De~Luca},\ and\ \citenamefont {Umeton}}]{peccianti_all-optical_2002}%
  \BibitemOpen
  \bibfield  {author} {\bibinfo {author} {\bibfnamefont {Marco}\ \bibnamefont {Peccianti}}, \bibinfo {author} {\bibfnamefont {Claudio}\ \bibnamefont {Conti}}, \bibinfo {author} {\bibfnamefont {Gaetano}\ \bibnamefont {Assanto}}, \bibinfo {author} {\bibfnamefont {Antonio}\ \bibnamefont {De~Luca}}, \ and\ \bibinfo {author} {\bibfnamefont {Cesare}\ \bibnamefont {Umeton}},\ }\bibfield  {title} {\enquote {\bibinfo {title} {All-optical switching and logic gating with spatial solitons in liquid crystals},}\ }\href {\doibase 10.1063/1.1519101} {\bibfield  {journal} {\bibinfo  {journal} {Applied Physics Letters}\ }\textbf {\bibinfo {volume} {81}},\ \bibinfo {pages} {3335--3337} (\bibinfo {year} {2002})}\BibitemShut {NoStop}%
\bibitem [{\citenamefont {Jakubowski}\ \emph {et~al.}(1997)\citenamefont {Jakubowski}, \citenamefont {Steiglitz},\ and\ \citenamefont {Squier}}]{jakubowski_information_1997}%
  \BibitemOpen
  \bibfield  {author} {\bibinfo {author} {\bibfnamefont {Mariusz~H.}\ \bibnamefont {Jakubowski}}, \bibinfo {author} {\bibfnamefont {Ken}\ \bibnamefont {Steiglitz}}, \ and\ \bibinfo {author} {\bibfnamefont {Richard}\ \bibnamefont {Squier}},\ }\bibfield  {title} {\enquote {\bibinfo {title} {Information transfer between solitary waves in the saturable {Schr{\"o}dinger} equation},}\ }\href {\doibase 10.1103/PhysRevE.56.7267} {\bibfield  {journal} {\bibinfo  {journal} {Physical Review E}\ }\textbf {\bibinfo {volume} {56}},\ \bibinfo {pages} {7267--7272} (\bibinfo {year} {1997})}\BibitemShut {NoStop}%
\bibitem [{\citenamefont {Sahoo}\ \emph {et~al.}(2022)\citenamefont {Sahoo}, \citenamefont {Mahato}, \citenamefont {Govindarajan},\ and\ \citenamefont {Sarma}}]{sahoo_bistable_2022}%
  \BibitemOpen
  \bibfield  {author} {\bibinfo {author} {\bibfnamefont {Ambaresh}\ \bibnamefont {Sahoo}}, \bibinfo {author} {\bibfnamefont {Dipti~Kanika}\ \bibnamefont {Mahato}}, \bibinfo {author} {\bibfnamefont {A.}~\bibnamefont {Govindarajan}}, \ and\ \bibinfo {author} {\bibfnamefont {Amarendra~K.}\ \bibnamefont {Sarma}},\ }\bibfield  {title} {\enquote {\bibinfo {title} {Bistable soliton switching dynamics in a {PT}-symmetric coupler with saturable nonlinearity},}\ }\href {\doibase 10.1103/PhysRevA.105.063503} {\bibfield  {journal} {\bibinfo  {journal} {Physical Review A}\ }\textbf {\bibinfo {volume} {105}},\ \bibinfo {pages} {063503} (\bibinfo {year} {2022})}\BibitemShut {NoStop}%
\bibitem [{\citenamefont {Umarov}\ \emph {et~al.}(2013)\citenamefont {Umarov}, \citenamefont {Messikh}, \citenamefont {Regaa},\ and\ \citenamefont {Baizakov}}]{umarov_variational_2013}%
  \BibitemOpen
  \bibfield  {author} {\bibinfo {author} {\bibfnamefont {B~A}\ \bibnamefont {Umarov}}, \bibinfo {author} {\bibfnamefont {A}~\bibnamefont {Messikh}}, \bibinfo {author} {\bibfnamefont {N}~\bibnamefont {Regaa}}, \ and\ \bibinfo {author} {\bibfnamefont {B~B}\ \bibnamefont {Baizakov}},\ }\bibfield  {title} {\enquote {\bibinfo {title} {Variational analysis of soliton scattering by external potentials},}\ }\href {\doibase 10.1088/1742-6596/435/1/012024} {\bibfield  {journal} {\bibinfo  {journal} {Journal of Physics: Conference Series}\ }\textbf {\bibinfo {volume} {435}},\ \bibinfo {pages} {012024} (\bibinfo {year} {2013})}\BibitemShut {NoStop}%
\bibitem [{\citenamefont {Morales-Molina}\ and\ \citenamefont {Vicencio}(2006)}]{morales-molina_trapping_2006}%
  \BibitemOpen
  \bibfield  {author} {\bibinfo {author} {\bibfnamefont {L.}~\bibnamefont {Morales-Molina}}\ and\ \bibinfo {author} {\bibfnamefont {R.A.}\ \bibnamefont {Vicencio}},\ }\bibfield  {title} {\enquote {\bibinfo {title} {Trapping of discrete solitons by defects in nonlinear waveguide arrays},}\ }\href {\doibase 10.1364/OL.31.000966} {\bibfield  {journal} {\bibinfo  {journal} {Optics Letters}\ }\textbf {\bibinfo {volume} {31}},\ \bibinfo {pages} {966--968} (\bibinfo {year} {2006})}\BibitemShut {NoStop}%
\bibitem [{\citenamefont {Al~Khawaja}\ \emph {et~al.}(2013)\citenamefont {Al~Khawaja}, \citenamefont {Al-Marzoug}, \citenamefont {Bahlouli},\ and\ \citenamefont {Kivshar}}]{al_khawaja_unidirectional_2013}%
  \BibitemOpen
  \bibfield  {author} {\bibinfo {author} {\bibfnamefont {U.}~\bibnamefont {Al~Khawaja}}, \bibinfo {author} {\bibfnamefont {S.~M.}\ \bibnamefont {Al-Marzoug}}, \bibinfo {author} {\bibfnamefont {H.}~\bibnamefont {Bahlouli}}, \ and\ \bibinfo {author} {\bibfnamefont {Yuri~S.}\ \bibnamefont {Kivshar}},\ }\bibfield  {title} {\enquote {\bibinfo {title} {Unidirectional soliton flows in {PT}-symmetric potentials},}\ }\href {\doibase 10.1103/physreva.88.023830} {\bibfield  {journal} {\bibinfo  {journal} {Physical Review A}\ }\textbf {\bibinfo {volume} {88}},\ \bibinfo {pages} {023830} (\bibinfo {year} {2013})}\BibitemShut {NoStop}%
\bibitem [{\citenamefont {Al~Khawaja}(2021)}]{al_khawaja_critical_2021}%
  \BibitemOpen
  \bibfield  {author} {\bibinfo {author} {\bibfnamefont {U.}~\bibnamefont {Al~Khawaja}},\ }\bibfield  {title} {\enquote {\bibinfo {title} {Critical soliton speed for quantum reflection by a reflectionless potential well},}\ }\href {\doibase 10.1103/PhysRevE.103.062202} {\bibfield  {journal} {\bibinfo  {journal} {Physical Review E}\ }\textbf {\bibinfo {volume} {103}},\ \bibinfo {pages} {062202} (\bibinfo {year} {2021})}\BibitemShut {NoStop}%
\bibitem [{\citenamefont {Javed}\ \emph {et~al.}(2021)\citenamefont {Javed}, \citenamefont {Uthayakumar}, \citenamefont {Alotaibi}, \citenamefont {Al-Marzoug}, \citenamefont {Bahlouli},\ and\ \citenamefont {Khawaja}}]{javed_unidirectional_2021}%
  \BibitemOpen
  \bibfield  {author} {\bibinfo {author} {\bibfnamefont {Amaria}\ \bibnamefont {Javed}}, \bibinfo {author} {\bibfnamefont {T.}~\bibnamefont {Uthayakumar}}, \bibinfo {author} {\bibfnamefont {M.~O.~D.}\ \bibnamefont {Alotaibi}}, \bibinfo {author} {\bibfnamefont {S.~M.}\ \bibnamefont {Al-Marzoug}}, \bibinfo {author} {\bibfnamefont {H.}~\bibnamefont {Bahlouli}}, \ and\ \bibinfo {author} {\bibfnamefont {U.~Al}\ \bibnamefont {Khawaja}},\ }\bibfield  {title} {\enquote {\bibinfo {title} {Unidirectional flow of composite bright-bright solitons through asymmetric double potential barriers and wells},}\ }\href {\doibase 10.1016/j.cnsns.2021.105968} {\bibfield  {journal} {\bibinfo  {journal} {Communications in Nonlinear Science and Numerical Simulation}\ }\textbf {\bibinfo {volume} {103}},\ \bibinfo {pages} {105968} (\bibinfo {year} {2021})}\BibitemShut {NoStop}%
\bibitem [{\citenamefont {Javed}\ \emph {et~al.}(2022{\natexlab{a}})\citenamefont {Javed}, \citenamefont {Susanto}, \citenamefont {Kusdiantara},\ and\ \citenamefont {Kourakis}}]{javed_unidirectional_2022}%
  \BibitemOpen
  \bibfield  {author} {\bibinfo {author} {\bibfnamefont {Amaria}\ \bibnamefont {Javed}}, \bibinfo {author} {\bibfnamefont {H.}~\bibnamefont {Susanto}}, \bibinfo {author} {\bibfnamefont {Rudy}\ \bibnamefont {Kusdiantara}}, \ and\ \bibinfo {author} {\bibfnamefont {I.}~\bibnamefont {Kourakis}},\ }\bibfield  {title} {\enquote {\bibinfo {title} {Unidirectional flow of symbiotic solitons and nonlinear modes of the {Schr{\"o}dinger} equation with an external potential},}\ }\href {\doibase 10.1140/epjp/s13360-022-03350-x} {\bibfield  {journal} {\bibinfo  {journal} {The European Physical Journal Plus}\ }\textbf {\bibinfo {volume} {137}},\ \bibinfo {pages} {1170} (\bibinfo {year} {2022}{\natexlab{a}})}\BibitemShut {NoStop}%
\bibitem [{\citenamefont {Sakkaf}\ and\ \citenamefont {Khawaja}(2023)}]{sakkaf_reflectionless_2023}%
  \BibitemOpen
  \bibfield  {author} {\bibinfo {author} {\bibfnamefont {L.~Al}\ \bibnamefont {Sakkaf}}\ and\ \bibinfo {author} {\bibfnamefont {U.~Al}\ \bibnamefont {Khawaja}},\ }\bibfield  {title} {\enquote {\bibinfo {title} {Reflectionless potentials and resonant scattering of flat-top and thin-top solitons},}\ }\href {\doibase 10.1103/PhysRevE.107.014202} {\bibfield  {journal} {\bibinfo  {journal} {Physical Review E}\ }\textbf {\bibinfo {volume} {107}},\ \bibinfo {pages} {014202} (\bibinfo {year} {2023})}\BibitemShut {NoStop}%
\bibitem [{\citenamefont {Kr{\'o}likowski}\ and\ \citenamefont {Kivshar}(1996)}]{krolikowski_soliton-based_1996}%
  \BibitemOpen
  \bibfield  {author} {\bibinfo {author} {\bibfnamefont {W.}~\bibnamefont {Kr{\'o}likowski}}\ and\ \bibinfo {author} {\bibfnamefont {Y.S.}\ \bibnamefont {Kivshar}},\ }\bibfield  {title} {\enquote {\bibinfo {title} {Soliton-based optical switching in waveguide arrays},}\ }\href {\doibase 10.1364/JOSAB.13.000876} {\bibfield  {journal} {\bibinfo  {journal} {Journal of the Optical Society of America B: Optical Physics}\ }\textbf {\bibinfo {volume} {13}},\ \bibinfo {pages} {876--887} (\bibinfo {year} {1996})}\BibitemShut {NoStop}%
\bibitem [{\citenamefont {{Asad-uz-zaman}}\ and\ \citenamefont {{Al Khawaja}}(2013)}]{asad-uz-zaman_directional_2013}%
  \BibitemOpen
  \bibfield  {author} {\bibinfo {author} {\bibfnamefont {M.}~\bibnamefont {{Asad-uz-zaman}}}\ and\ \bibinfo {author} {\bibfnamefont {U.}~\bibnamefont {{Al Khawaja}}},\ }\bibfield  {title} {\enquote {\bibinfo {title} {Directional flow of solitons with asymmetric potential wells: {Soliton} diode},}\ }\href {\doibase 10.1209/0295-5075/101/50008} {\bibfield  {journal} {\bibinfo  {journal} {EPL (Europhysics Letters)}\ }\textbf {\bibinfo {volume} {101}},\ \bibinfo {pages} {50008} (\bibinfo {year} {2013})}\BibitemShut {NoStop}%
\bibitem [{\citenamefont {Al~Khawaja}\ and\ \citenamefont {Sukhorukov}(2015)}]{al_khawaja_unidirectional_2015}%
  \BibitemOpen
  \bibfield  {author} {\bibinfo {author} {\bibfnamefont {U.}~\bibnamefont {Al~Khawaja}}\ and\ \bibinfo {author} {\bibfnamefont {Andrey~A.}\ \bibnamefont {Sukhorukov}},\ }\bibfield  {title} {\enquote {\bibinfo {title} {Unidirectional flow of discrete solitons in waveguide arrays},}\ }\href {\doibase 10.1364/OL.40.002719} {\bibfield  {journal} {\bibinfo  {journal} {Optics Letters}\ }\textbf {\bibinfo {volume} {40}},\ \bibinfo {pages} {2719} (\bibinfo {year} {2015})}\BibitemShut {NoStop}%
\bibitem [{\citenamefont {Javed}\ \emph {et~al.}(2022{\natexlab{b}})\citenamefont {Javed}, \citenamefont {Uthayakumar},\ and\ \citenamefont {Al~Khawaja}}]{javed_simulating_2022}%
  \BibitemOpen
  \bibfield  {author} {\bibinfo {author} {\bibfnamefont {Amaria}\ \bibnamefont {Javed}}, \bibinfo {author} {\bibfnamefont {T.}~\bibnamefont {Uthayakumar}}, \ and\ \bibinfo {author} {\bibfnamefont {U.}~\bibnamefont {Al~Khawaja}},\ }\bibfield  {title} {\enquote {\bibinfo {title} {Simulating an all-optical quantum controlled-{NOT} gate using soliton scattering by a reflectionless potential well},}\ }\href {\doibase 10.1016/j.physleta.2022.127949} {\bibfield  {journal} {\bibinfo  {journal} {Physics Letters A}\ }\textbf {\bibinfo {volume} {429}},\ \bibinfo {pages} {127949} (\bibinfo {year} {2022}{\natexlab{b}})}\BibitemShut {NoStop}%
\bibitem [{\citenamefont {Kivshar}\ and\ \citenamefont {Agrawal}(2003)}]{kivshar_optical_2003}%
  \BibitemOpen
  \bibfield  {author} {\bibinfo {author} {\bibfnamefont {Yuri~S.}\ \bibnamefont {Kivshar}}\ and\ \bibinfo {author} {\bibfnamefont {Govind~P.}\ \bibnamefont {Agrawal}},\ }\href@noop {} {\emph {\bibinfo {title} {Optical {Solitons}: {From} {Fibers} to {Photonic} {Crystals}}}}\ (\bibinfo  {publisher} {Academic Press},\ \bibinfo {year} {2003})\BibitemShut {NoStop}%
\bibitem [{\citenamefont {Adhikari}(2005)}]{adhikari_bright_2005}%
  \BibitemOpen
  \bibfield  {author} {\bibinfo {author} {\bibfnamefont {Sadhan~K.}\ \bibnamefont {Adhikari}},\ }\bibfield  {title} {\enquote {\bibinfo {title} {Bright solitons in coupled defocusing {NLS} equation supported by coupling: {Application} to {Bose}--{Einstein} condensation},}\ }\href {\doibase 10.1016/j.physleta.2005.07.044} {\bibfield  {journal} {\bibinfo  {journal} {Physics Letters A}\ }\textbf {\bibinfo {volume} {346}},\ \bibinfo {pages} {179--185} (\bibinfo {year} {2005})}\BibitemShut {NoStop}%
\bibitem [{\citenamefont {P{\"o}schl}\ and\ \citenamefont {Teller}(1933)}]{poschl_bemerkungen_1933}%
  \BibitemOpen
  \bibfield  {author} {\bibinfo {author} {\bibfnamefont {G.}~\bibnamefont {P{\"o}schl}}\ and\ \bibinfo {author} {\bibfnamefont {E.}~\bibnamefont {Teller}},\ }\bibfield  {title} {\enquote {\bibinfo {title} {Bemerkungen zur {Quantenmechanik} des anharmonischen {Oszillators}},}\ }\href {\doibase 10.1007/BF01331132} {\bibfield  {journal} {\bibinfo  {journal} {Zeitschrift f{\"u}r Physik}\ }\textbf {\bibinfo {volume} {83}},\ \bibinfo {pages} {143--151} (\bibinfo {year} {1933})}\BibitemShut {NoStop}%
\bibitem [{\citenamefont {Neshev}\ and\ \citenamefont {Aharonovich}(2018)}]{Neshev2018}%
  \BibitemOpen
  \bibfield  {author} {\bibinfo {author} {\bibfnamefont {Dragomir}\ \bibnamefont {Neshev}}\ and\ \bibinfo {author} {\bibfnamefont {Igor}\ \bibnamefont {Aharonovich}},\ }\bibfield  {title} {\enquote {\bibinfo {title} {Active meta-optics and nanophotonics with halide perovskites},}\ }\href {\doibase 10.1038/s41377-018-0058-1} {\bibfield  {journal} {\bibinfo  {journal} {Light: Science \& Applications}\ }\textbf {\bibinfo {volume} {7}},\ \bibinfo {pages} {58} (\bibinfo {year} {2018})}\BibitemShut {NoStop}%
\end{thebibliography}%

\end{document}